\documentclass[traditabstract]{aa} 

\usepackage{subfig}
\usepackage{color}
\usepackage{natbib}
\usepackage{url}
\bibpunct{(}{)}{;}{a}{}{,} 
\usepackage{graphicx,latexsym,amssymb}

\begin{document}
\newcommand\arcdeg{^{\circ}}
\newcommand\arcs{^{\prime\prime}}
\newcommand{\arcm}{^\prime}
\newcommand{\fermi}{\textit{Fermi}/LAT}
\newcommand{\hb}{\object{HB\,21}}

\title{An extended source of GeV gamma rays coincident with the supernova remnant \hb}
\subtitle{}
\titlerunning{GeV gamma rays coincident with \hb}

\author{Ignasi Reichardt\inst{1}
\and	Emma de O\~na-Wilhelmi\inst{2,3}
\and    Javier Rico\inst{1,4}
\and	Rui-zhi Yang\inst{2}
}

\institute{
\ IFAE, Edifici C7b, Campus UAB, E-08193 Bellaterra, Spain
\and Max-Planck-Institut f{\"u}r Kernphysik, P.O. Box 103980, 69029 Heidelberg, Germany
\and now at Institut de Ci\`encies de l'Espai (IEEC-CSIC), E-08193 Bellaterra, Spain
\and Instituci{\'o} Catalana de Recerca i Estudis Avan\c{c}ats (ICREA), E-08010 Barcelona, Spain
}

\date{Received 4 July 2012 / Accepted 22 August 2012}
\offprints{I.~Reichardt (ignasi@ifae.cat)}

\begin{abstract}{
We analyze 3.5 years of public \fermi\, data around the position of the supernova remnant \hb, where four point-like sources from the $2^{nd}$ \fermi\, catalog are located. We determine that the gamma-ray source is produced by a single extended source. We model the observed morphology as a uniform circle. The spectral energy distribution is best described by a curved power law, with a maximum at $413\pm11$\,MeV. We divide the circle into three regions defined by previously identified shocked molecular clouds, and find that one of these regions has a softer spectrum. The $>3$\,GeV gamma-ray emission of the soft spectrum region is bow-shaped and coincident with the supernova remnant shell seen at radio wavelengths. We suggest that the gamma-ray emission from \hb\, can be understood as a combination of emission from shocked/illuminated molecular clouds, one of them coincident with the supernova remnant shell itself.}
\end{abstract}
  

\keywords{Acceleration of particles - cosmic rays - ISM: supernova remnants - ISM: clouds - Gamma rays: general - Gamma rays: ISM}
\maketitle

\section{Introduction}
\label{intro}
Gamma-ray emission from SNRs is key for understanding the processes that lead to particle acceleration in the Galaxy and, in particular, to the production of cosmic rays. Since protons cannot be traced back to distant sources, proton acceleration in SNRs may be probed by observing their effects on dense molecular clouds located in the vicinity of the accelerating region \citep{Gabici2009}. In particular, the observation of gamma rays from neutral pion decay is considered to be the smoking gun of cosmic-ray production, since they trace the collisions of accelerated protons with nucleons from the ambient medium. Examples of such interaction are firmly established in cases like that of the SNR W44 \citep{w44}, where GeV gamma rays emanate from regions clearly offset from the SNR shell.

\hb\, (G89.0+4.7) is a 19000 year old\footnote{We quote here the commonly accepted age for this object, but we note that there are indications that \hb\, could have an age of the order of 5000 years \citep{Lazendic2006}} \citep{Leahy1996} mixed-morphology supernova remnant (SNR) at a distance of 0.8\,kpc \citep{tatematsu}. As seen in radio continuum images, the SNR displays an elliptical shell of $2^\circ\times1.5^\circ$ \citep{condon} (mean diameter of $\sim25$\,pc), slightly tilted in the NW-SE direction. Only weak, center-filling X-ray emission of thermal origin is associated with \hb \citep{Lazendic2006}.

The interaction of \hb\, with the surrounding interstellar medium (ISM) has been intensively studied. Given the absence of OH masers around \hb\, \citep{frail}, the evidence of interaction between the blast wave and the ISM is established by means of local dynamic effects that broaden emission lines. In the following paragraph we briefly describe the clouds that present such broad emission lines. We refer to Fig. 1 in \citet{byun} for a detailed view of the interacting clouds, or Fig.~\ref{signimaps} in this work for a schematic view.

Evidence of shocked molecular gas was found by \citet{koo} in the northern and in the southern parts of the shell. The northern cloud (cloud N hereafter) consists of several small, bright clumps plus a diffuse component extending to the E. The southern cloud (cloud S) presents a complex filamentary structure, with a velocity spread of up to 40\,km~s$^{-1}$ for some particular clumps, and it is coincident with a mass of shocked atomic gas detected by \citet{kooheiles1991}. There is also a bow-shaped cloud at the NW rim of the radio shell (cloud NW) \citep{byun}, and the central thermal X-ray bright area is occupied by small evaporating clouds. Moreover the so-called clouds A, B, and C \citep{tatematsu} are aligned N-S in the approximately straight E rim of the SNR. These clouds may be regarded as overdensities of the giant molecular cloud of the Cyg OB7 association \citep{HuangThaddeus1986}, which provides the distance estimate for \hb. Clouds A, B, and C are located where the eastward blast wave apparently collides with the so-called \textit{wall}. The wall consists of a sharp edge of otherwise smoothly distributed atomic gas, which extends beyond the SNR boundary, both N and S, therefore suggesting that it is a pre-existing structure that affects the evolution of the SNR and not the other way around. Probably, the wall is the border of the cavity resulting from a former HII region around the \hb\, progenitor, which might be a former member of the Cyg OB7 association \citep{tatematsu}. There is also the possibility that the coincidence of the A, B, C clouds and the wall with the SNR shell is a projection effect, where the clouds are indeed in the vicinity of the SNR, but they lie in front or behind it \citep{koo}. \citet{byun} suggested that the SNR could be as far as 1.7\,kpc, in which case the whole Cyg OB7 complex would be in the foreground.

According to~\cite{fermicat} three point-like sources (2FGL J2041.5+5003, 2FGL J2043.3+5105 and 2FGL J2046.0+4954) in the second \fermi\, source catalog (2FGL catalog hereafter) are coincident with the extended radio emission of \hb. In this article we report a detailed analysis of the public \fermi\, data that will lead to a deeper understanding of the gamma-ray emission from this object.

\section{Data analysis}
\label{analysis}
We analyzed \fermi\, data with the LAT analysis software, the \textit{ScienceTools} version v9r27p1\footnote{data and software are publicly available at \url{http://fermi.gsfc.nasa.gov/ssc}}. We analyzed Pass-7 data corresponding to the period between August $4^{th}$ 2008 (start of science operations) and February 2, 2012. Since \hb\, is almost $5\arcdeg$ off the galactic plane, it allows for a notably background-reduced analysis with respect to the sources found at lower galactic latitudes. To exploit this advantage, we defined a region of interest (ROI, i.e. the sky region whose LAT photon events are considered) as a circle of $10\arcdeg$ radius centered on the position ($\alpha, \delta$) = ($20^{h}41^{m}05^{s}$, $51\arcdeg15\arcm58\arcs$), which is displaced by $1\arcdeg$ toward positive galactic latitudes with respect to the catalog position of \hb \citep{census}. We selected class 2 events in the energy range between 100\,MeV and 100\,GeV. We applied a set of quality cuts, including the requirement for the spacecraft to be in normal operation mode (LAT\_CONFIG==1), data to be flagged as good quality (DATA\_QUAL==1), and a cut on the rocking angle of the spacecraft (ABS(ROCK\_ANGLE)$<52\arcdeg$). In addition, we applied a zenith angle cut of $100\arcdeg$ in order to prevent event contamination from the Earth limb. Data are binned in sky coordinates with the \texttt{gtbin} tool, using square bins of $0.125\arcdeg$ side. We refer later to the two-dimensional histograms resulting from \texttt{gtbin} as \textit{count maps}.

To study the morphological and spectral properties of \hb, we performed a three-dimensional (two spatial dimensions plus the energy) maximum likelihood analysis, using the standard \texttt{gtlike} tool. In this method, the likelihood is computed for different models defined by the position and morphology of the sources producing gamma rays in the ROI. For each source, a different spectral shape is assumed, and the spectral parameters are left free in the likelihood maximization. Our starting point consists of the standard galactic and extragalactic diffuse emission models provided in the \textit{ScienceTools}, plus the point-like sources in the 2FGL catalog lying up to $15\arcdeg$ away of the ROI center, excluding 2FGL J2041.5+5003, 2FGL J2043.3+5105, 2FGL J2046.0+4954 and 2FGL J2051.8+5054. We note that according to \cite{fermicat}, the source 2FGL J2051.8+5054 is not associated to \hb, but it lies very close to the NE edge of the SNR shell, in remarkable coincidence with the above-mentioned cloud A (see Fig.~\ref{100MeV}). For this reason, in the following section we consider 2FGL J2051.8+5054 as part of the gamma-ray emission related to \hb. We refer to this model as the \textit{null hypothesis}, since it assumes that the gamma-ray emission from \hb\, is faint enough to not be distinguished from the background. Next, we compare the maximum likelihood obtained with the null hypothesis to those of several models that include various morphological descriptions of the GeV emission from \hb. First, we include the four point-like sources mentioned above. In addition, we explore the possibility that the observed emission is from an extended, resolved source. For this, the four point-like sources associated to \hb\, are replaced by extended source templates. The scenarios considered are detailed in sect.~\ref{morphology}. For each model, the goodness of the likelihood fit is estimated by means of a test statistic (TS) defined as
\begin{equation}
  \rm{TS} = -2\log(\mathcal{L}_0/\mathcal{L}),
\end{equation}
where $\mathcal{L}_{0}$ and $\mathcal{L}$ are the likelihood of the null hypothesis and the tested models respectively. Since TS is a likelihood ratio of two nested models, it asymptotically follows a $\chi^2$ distribution with a number of degrees of freedom equal to the extra number of parameters of the test hypothesis with respect to the null one \citep{brightlist}. Despite some caveats \citep{Protassov}, it is normally accepted TS = 25 as a detection threshold of a source with two spectral parameters (flux normalization and spectral index), which corresponds to a statistical significance of 4.6 sigma. In all cases, \texttt{gtlike} is run in a two-step procedure: first allowing a loose tolerance up to 10\% in the fit parameters and using the \texttt{MINUIT} method and, second, using the output of \texttt{MINUIT} as the initial value for a refitting of the model parameters  with a tighter requirement of 0.1\% tolerance and using the \texttt{NEWMINUIT} method. The output of \texttt{gtlike} allows us to generate \textit{synthetic maps} with the expected source shape and brightness, given the model best-fit parameters. These synthetic maps can be subtracted from the counts maps in order to visualize the disagreements between the real data and its parametrization, in the form of a \textit{residuals map}. We then divide, pixel by pixel, the residuals map by the square root of the number of counts in the synthetic map, thus obtaining a measure of the significance of the disagreement in every pixel, which we will call the \textit{signal-to-noise ratio, S/N, map}.

\section{Results}
\label{results}
\subsection{Morphology}
\label{morphology}
\begin{figure*}
   \centering
   \subfloat[S/N map above 100\,MeV. Red crosses are the point-like sources from 2FGL catalog surrounded by their 95\% position uncertainty ellipses.  Sources highlighted with green ellipse are those ones we consider related to \hb\, and are removed from the model when computing the null hypothesis map.]{\label{100MeV}\includegraphics[width=0.45\textwidth]{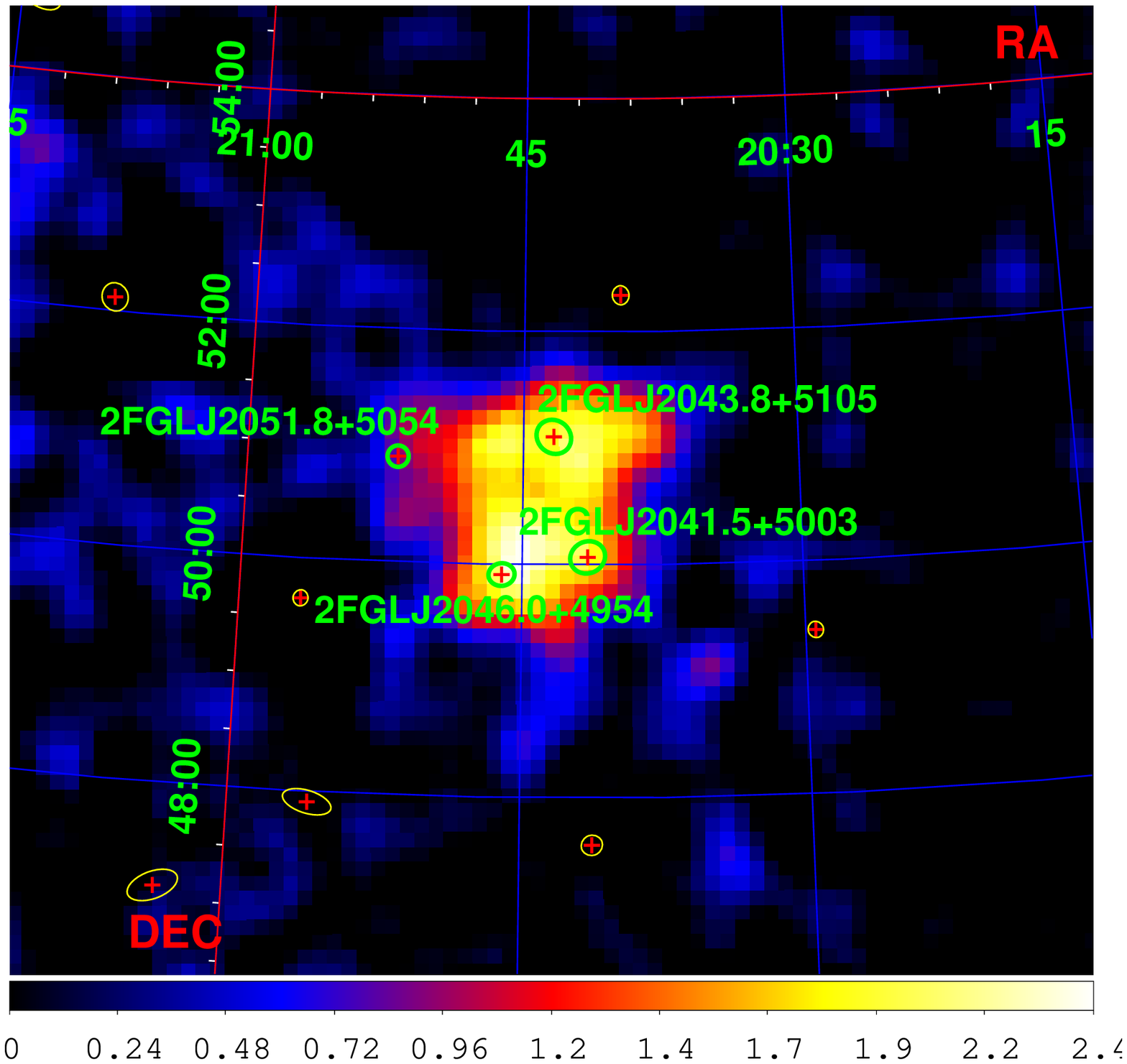}}
   ~
   \subfloat[S/N map above 500\,MeV. The circle used for the spectral modeling is shown, as well as the divisions discussed in table~\ref{partcircle}. Symbols A, B, C, NW, N, and S mark the position of the relevant clouds, mentioned in section~\ref{intro}. The size of the markers is not related to the extension of the clouds.]{\label{500MeV}\includegraphics[width=0.45\textwidth]{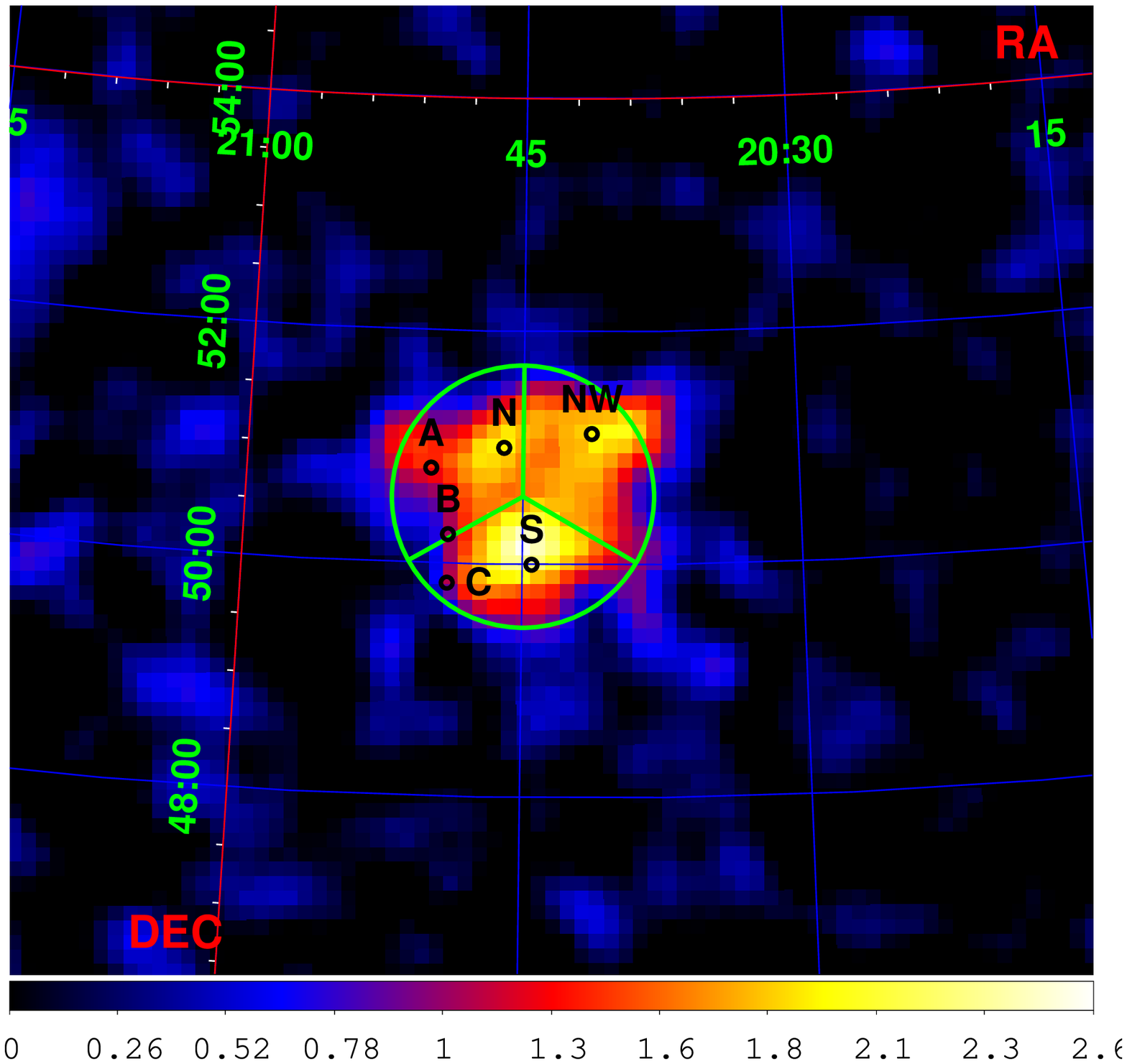}}

   \subfloat[S/N map above 1\,GeV.  White contours represent the large-scale CO distribution from \cite{dame2001} integrated between 0 and -20\,km~s$^{-1}$. Symbols A, B, C, NW, N, and S mark the position of the relevant clouds, mentioned in section~\ref{intro}.]{\label{1GeV}\includegraphics[width=0.45\textwidth]{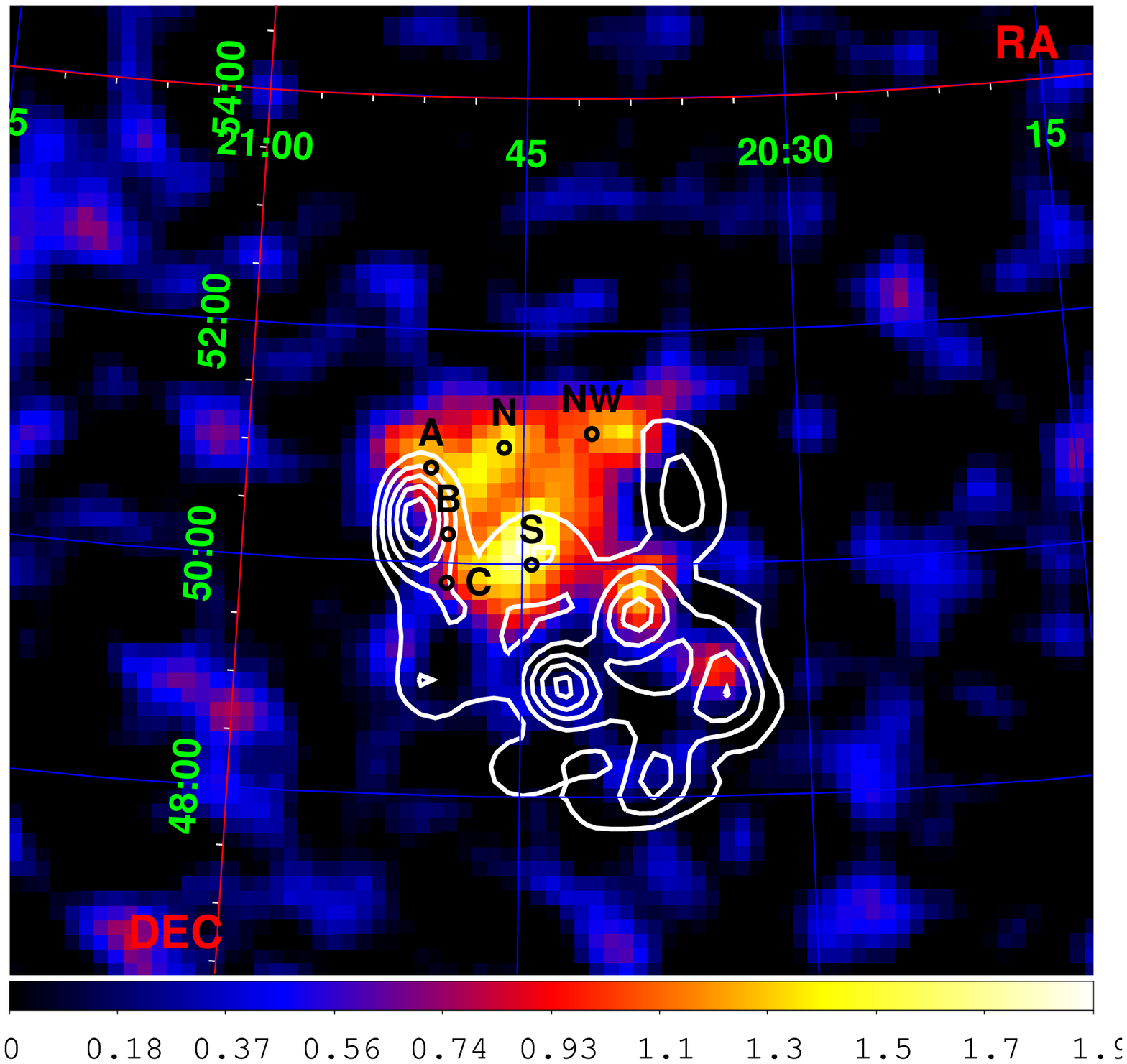}}
   ~
   \subfloat[S/N map above 3\,GeV. Green contours depict the 4850\,MHz radio continuum emission \citep{condon}.]{\label{3GeV}\includegraphics[width=0.45\textwidth]{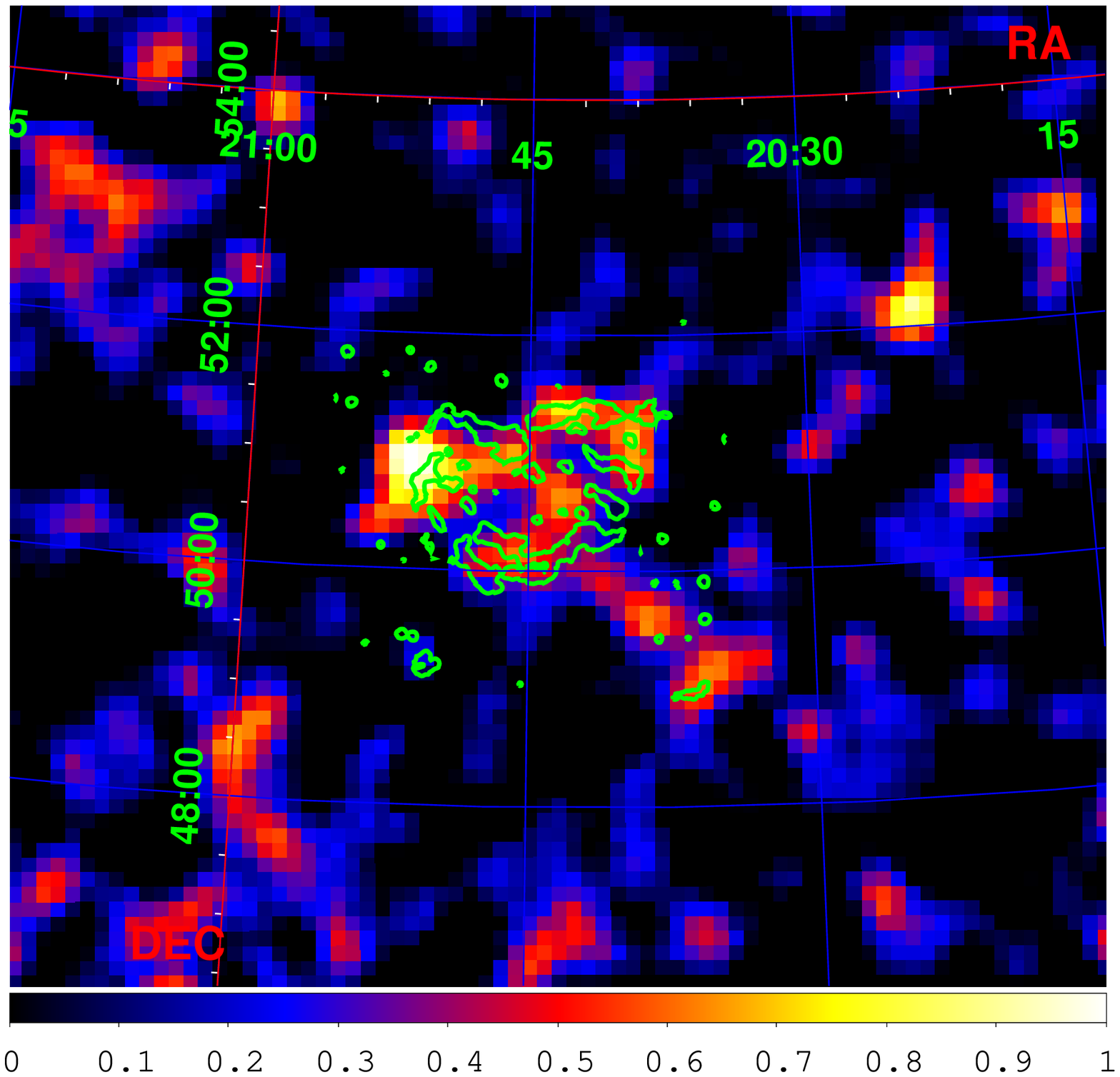}}
   \caption{S/N maps above 100\,MeV, 500\,MeV, 1\,GeV, and 3\,GeV. The color scale represents signal-to-noise ratio (defined as real counts minus model counts divided by square root of model counts) for the null hypothesis. Different multiwavelength information is included in each panel.}
\label{signimaps}
\end{figure*}

To visualize a possible gamma-ray source associated to \hb, we produce the S/N map corresponding to the null hypothesis, in different energy ranges (Fig.~\ref{signimaps}). The point spread function\footnote{measured as the 68\% event containment radius.} (PSF) of \fermi\, is up to $3\arcdeg$ at 100\,MeV energies, improves to $\sim0.4\arcdeg$ at 1\,GeV and achieves $\sim0.1\arcdeg$ above 10\,GeV \citep{fermiperformance}. Therefore, at the lowest energies (Fig.~\ref{100MeV}) it is not possible to distinguish any structure beyond a more or less flat emission extended throughout the SNR shell. Above 500\,MeV (Fig.~\ref{500MeV}) and above 1\,GeV (Fig.~\ref{1GeV}) a shell-like structure becomes visible.

Above 3\,GeV (Fig.~\ref{3GeV}) we can identify several structures. The most remarkable feature is the NW arc, which coincides with the SNR shell and the position of cloud NW. There is also a bright spot close to cloud A. This component seems to become more prominent with increasing energy. The center of the SNR does not show especially bright emission, whereas the S part of the shell presents an enhancement roughly coinciding with cloud S. We show below that the emission above 3\,GeV is still significant (Table~\ref{likelihood}).

To evaluate the morphological properties of the source, several scenarios were considered:
\begin{enumerate}
\item{The four point-like sources from 2FGL coincident with \hb;}
\item{The 4850MHz map from~\citep{condon}, where the quasar \object{3c418.0} was removed from the radio map (see discussion below);}
\item{A circle of flat emission centered on the catalog position of \hb.}
\end{enumerate}
All the templates were rebinned to match the field of view and the pixel size of the \fermi\, maps. In Table~\ref{likelihood} we show the TS values for each model in several energy regimes, as well as the number of additional parameters in each model. The point-like sources model introduces additional degrees of freedom for the flux normalization of each of the four sources, plus the spectral indices. 2FGL J2043.3+5105 and 2FGL J2046.0+4954 contain an additional parameter $\beta$ which allows an energy-dependent spectral index $-\alpha-\beta\log\left(E/1000\mathrm{GeV}\right)$. The spectrum of the circle and the 4850\,MHz templates are described by a power-law function at this stage.

\begin{table}
  \caption[]{TS of the spatial models above different minimum energies. $N$ is the number of additional parameters of the model with respect to the null hypothesis accounting for the spectral index and normalization factor of each additional source or component.}
         \label{likelihood}
         \begin{tabular}{llllll}
           \hline\hline
           Model  & N & $TS_{100}$ & $TS_{500}$ & $TS_{1000}$ & $TS_{3000}$ \\
           \hline
           1. 2FGL sources & 10 & 959 & 610 & 245 & 48 \\
           2. Circle       &  2 & 832 & 626 & 279 & 53 \\
           3. 4850\,MHz    &  2 & 780 & 613 & 275 & 42 \\
           \hline
         \end{tabular}
\end{table}

All of the models listed above provide a good description of the HB21 field compared to that of the null hypothesis. Thus, one or more sources are detected in the \hb\, region with high significance, even at energies above 3\,GeV. However, we discard the description by means of point-like sources, since it provides the highest TS only at the lowest energies (Table~\ref{likelihood}), where the broad PSF of \fermi\, does not allow disentangling any substructure. The model with the point-like sources introduces ten degrees of freedom. Therefore we consider that the improvement in TS at low energies comes from the greater number of parameters and not from the point-like source model being a better description of the morphology. With only two degrees of freedom, the overall emission can be described well at all energies by a circle of radius $1.125\arcdeg$ centered on the catalog position of the SNR. The radius of the circle is chosen to provide the highest TS after varying it from 0.75 to $1.375\arcdeg$ in steps of $0.125\arcdeg$ (one pixel). Initially, we assume a simple power-law function as spectral model. 

\begin{figure*}
\centering
\subfloat[$E>100$\,MeV]{\label{100MeVprof}\includegraphics[width=0.45\textwidth]{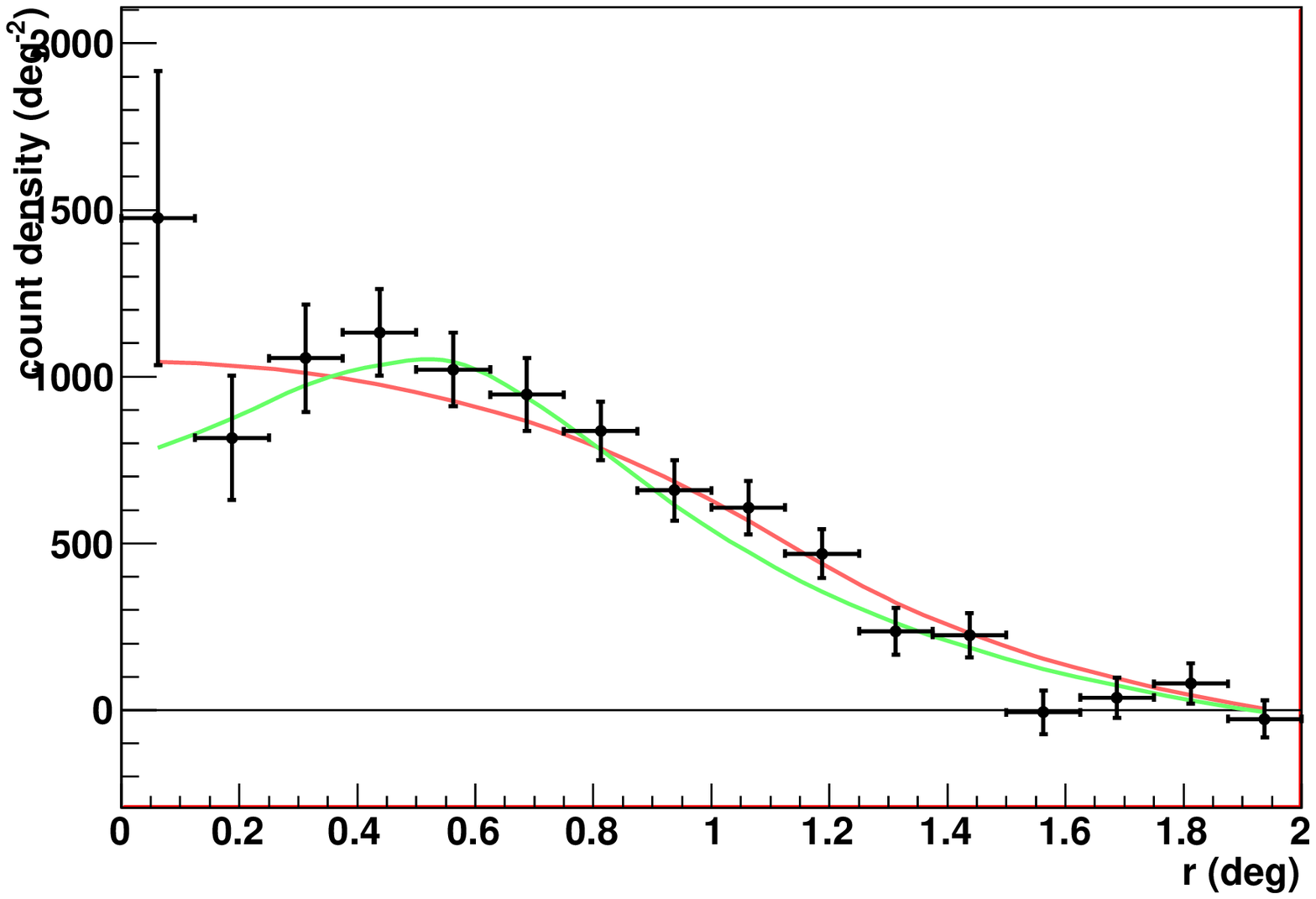}}
~
\subfloat[$E>500$\,MeV]{\label{500MeVprof}\includegraphics[width=0.45\textwidth]{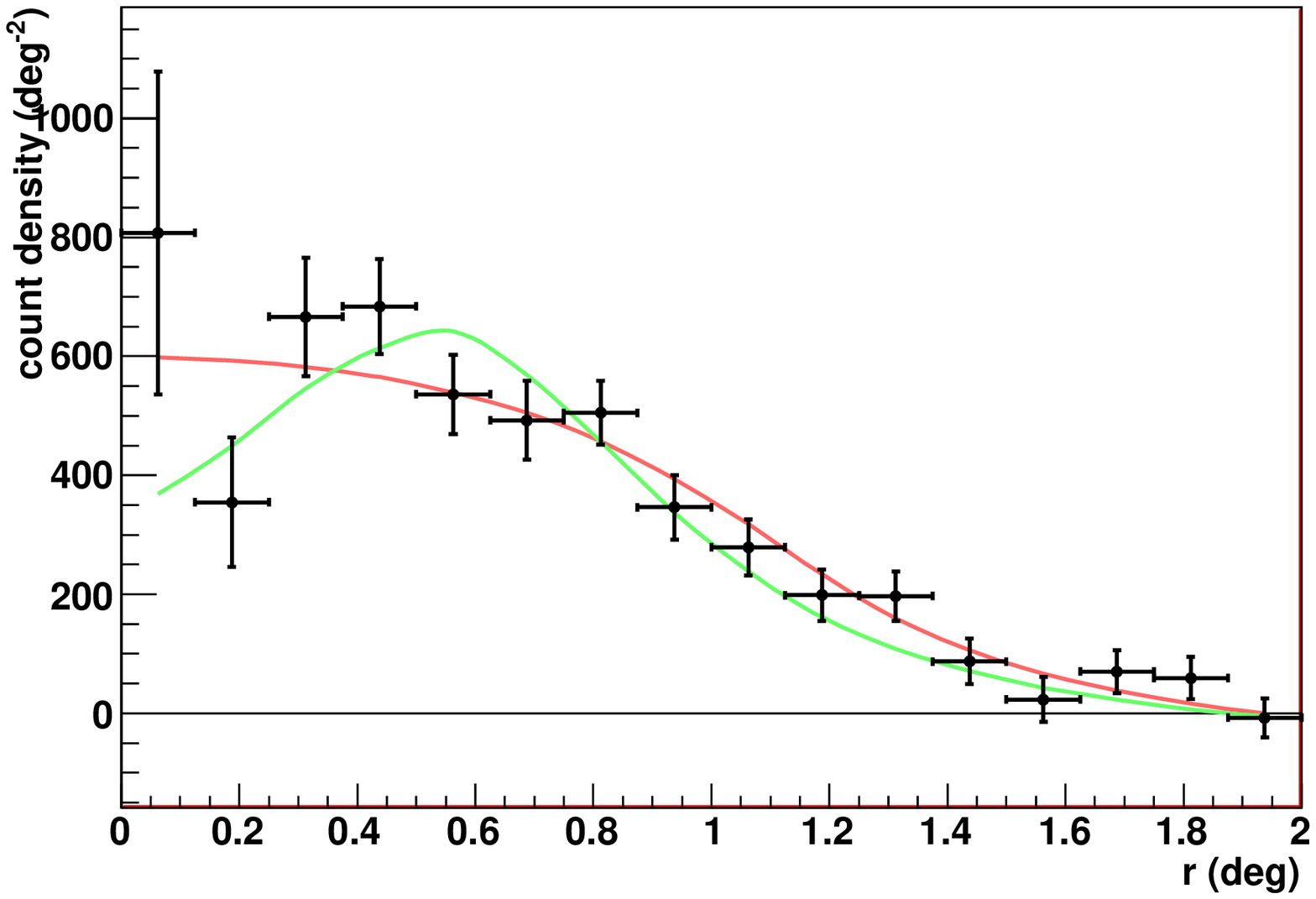}}

\subfloat[$E>1$\,GeV]{\label{1GeVprof}\includegraphics[width=0.45\textwidth]{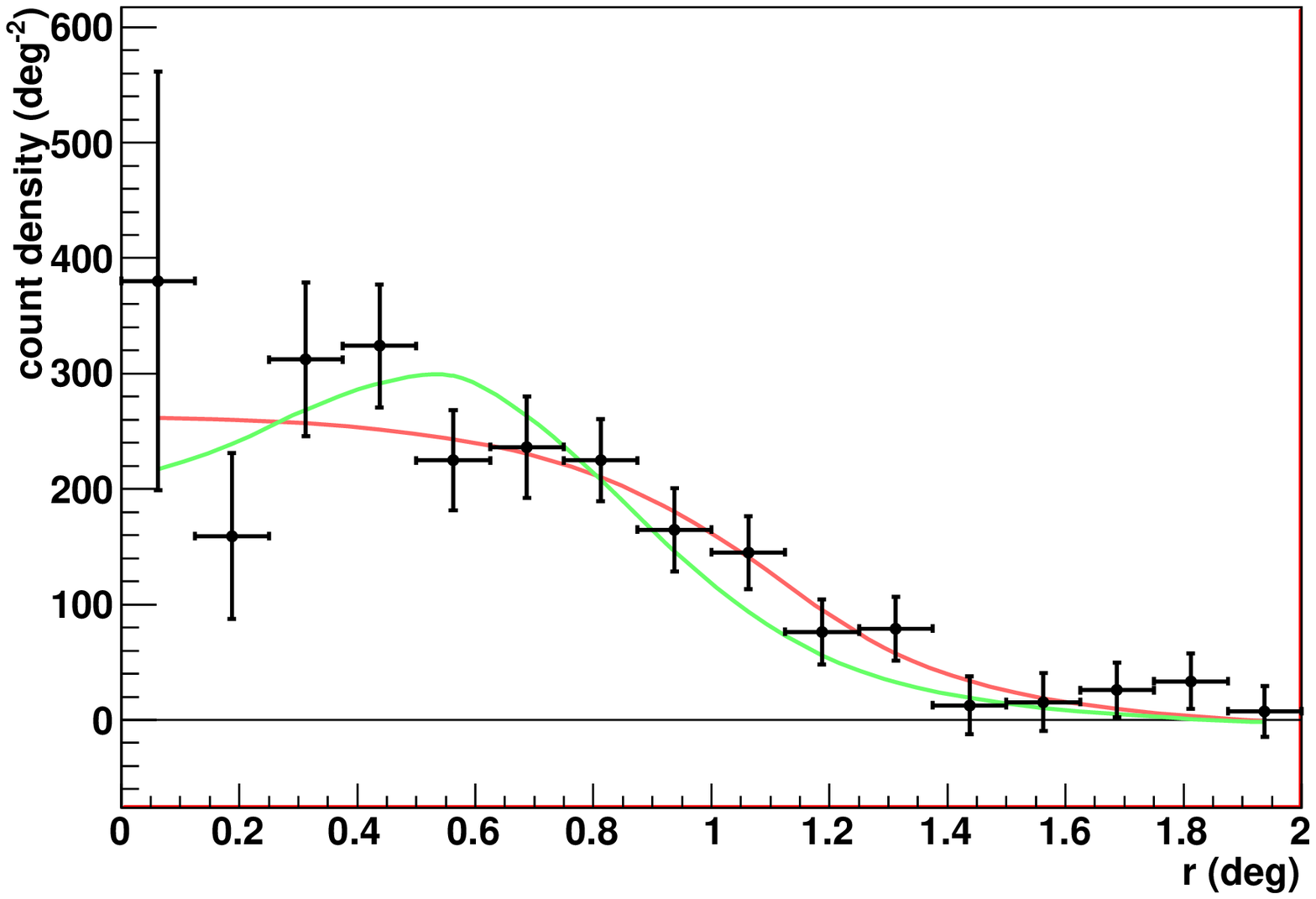}}
~
\subfloat[$E>3$\,GeV]{\label{3GeVprof}\includegraphics[width=0.45\textwidth]{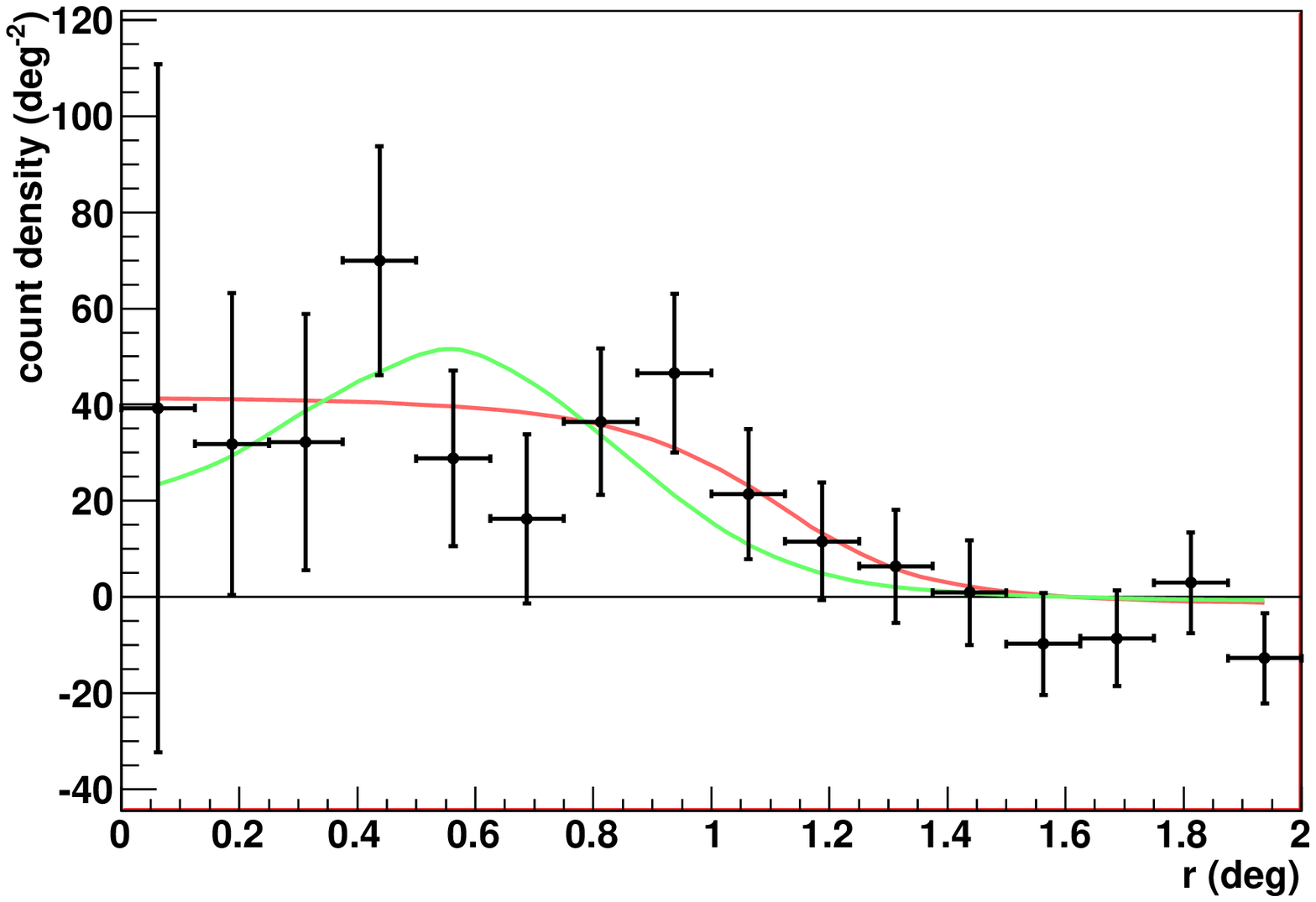}}
\caption{Radial profile of the excess (real counts minus model counts generated with the null hypothesis model). Red curve represents the profile expected from a flat circular emission region of $1.125\arcdeg$ radius. Green curve represents the profile expected from a gamma-ray emission following the radio continuum emission at 4850\,MHz.}
\label{profiles}
\end{figure*}

To further investigate the possibility of a shell-like morphology we produced the radial profiles of the excess (real counts minus model counts, for the null hypothesis) in each energy range, and we compared this profile with the one expected from models (2) and (3) (Fig.~\ref{profiles}). In view of the profiles, it is not conclusive whether the gamma-ray emission can be related to a shell-like structure. However, given the hints obtained so far, we considered this possibility by testing several ring-shaped models with various inner and outer radii. The outer radius is varied within the same range as for the optimization of the flat circle, whereas we considered inner radii ranging from 0 (circle) to $0.75\arcdeg$, also in steps of $0.125$. The best combination at all energies is $0.125\arcdeg$ for the inner radius and $1.125\arcdeg$ for the outer radius. However, the difference in TS is about 0.1 for all energies, which  is not significant. In addition, we consider that smearing by the PSF would prevent such a narrow hole (of only two pixels) from being distinguished if used, so we continued to use the flat circle for the subsequent analysis.

We addressed the possibility of having additional point-like sources besides the above-mentioned ones. First, the NW corner of the radio shell reveals a bright point-like radio source due to the presence of the quasar 3c418.0 at $z=1.6865$ \citep{redshift}. We considered the possibility that this quasar contributes to the gamma-ray emission. To do so, we looked for variability in the GeV signal by means of an aperture analysis within a radius of one degree around 3c418.0. After several temporal binnings we did not find any significant variability in the photon rate around the quasar. Therefore, we conclude that if this object has any contribution, we cannot disentangle it from that of the SNR with the current data set. Secondly, in the 1\,GeV map (Fig.~\ref{1GeV}) two spots appear SW of the SNR. Comparing with the CO large-scale distribution around the SNR we find that these spots are roughly coincident with local maxima of the gas distribution. Taking this possibility into account, we repeated the likelihood analysis adding two sources at the position of the gas overdensities. None of these sources were significant, whether in the analysis above 100\,MeV energies or above 1\,GeV.

\subsection{Spectral energy distribution}
\label{spectrum}
\begin{figure}
   \centering
   \includegraphics[width=0.45\textwidth]{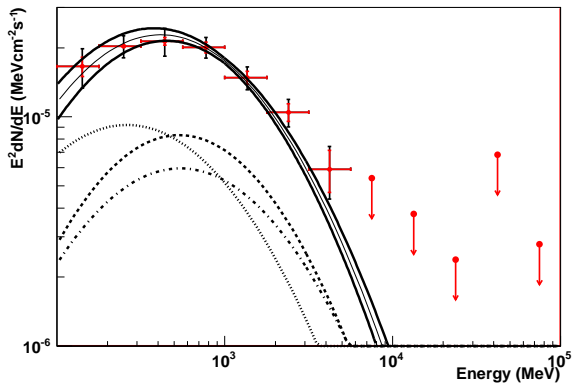}
   \caption{Spectral energy distribution of the gamma-ray emission from \hb\, modeled as a flat circle of 1.125$\arcdeg$ radius. Red error bar is statistical uncertainty. An additional systematic uncertainty of 10\% ($E<560$\,MeV) and of 5\% ($E>560$\,MeV) is represented by the black error bar. The solid thin black curve is the log-parabola (equation~\ref{logparabola}) used to model the overall spectrum. Curves with extreme values of $\alpha$ and $\beta$, within statistical uncertainty are also shown (solid thick black curves). Dotted, dashed, and dash-dotted curves are the best spectral descriptions for segments NW, S, and NE respectively (see Table~\ref{partcircle}).}
   \label{spec}
\end{figure}

We divide the considered energy range (100\,MeV to 100\,GeV) into twelve bins and compute the spectral energy distribution of the whole source by extracting its flux in each bin (Fig.~\ref{spec}). Only those bins with TS$>10$ ($\sim 3.2 \sigma$) are shown as spectral points. The last significant bin is the one from 3.2\,GeV to 5.6\,GeV. In addition we show 95\% confidence level upper limits for the explored energy range, up to 100\,GeV. The upper limits correspond to the flux providing a likelihood value such that $2\Delta\log\mathcal{L}=4$.

We notice that the spectrum deviates from a power-law function and suggests the presence of a peak at few hundred MeV. We test the possibility that the gamma ray emission is described by a smoothly broken power law of the form
\begin{equation}
\label{bpl}
  \frac{dN}{dE}=N_0\left(\frac{E}{100\mathrm{MeV}}\right)^{\gamma_1}\left(1+\left(\frac{E}{E_b}\right)^{\frac{\gamma_1-\gamma_2}{0.5}}\right)^{-0.5},
\end{equation}
or a curved power law (log-parabola)
\begin{equation}
  \frac{dN}{dE}=N_0\left(\frac{E}{1000\mathrm{MeV}}\right)^{-\alpha-\beta\log{\frac{E}{1000\mathrm{MeV}}}},
\label{logparabola}
\end{equation}
or a power law with a cut-off
\begin{equation}
   \frac{dN}{dE}=N_0\left(\frac{E}{1000\mathrm{MeV}}\right)^{-\gamma}\exp{\frac{E}{E_{cutoff}}}.
\label{plcutoff}
\end{equation}

We use the likelihood ratio $-2\log(\mathcal{L}_{pl}/\mathcal{L}_{model})$ as a measure of the improvement of the likelihood fit with respect to the simple power law, when different spectral shapes are used. We use equations \ref{bpl} (two extra degrees of freedom), \ref{logparabola}, and \ref{plcutoff} (one extra degree of freedom), resulting in likelihood ratios of 143, 146, and 130, respectively. Provided that the log-parabola introduces an additional parameter ($\beta$) to the simple power law, we conclude that the chance probability of the log-parabola being a better description of the spectrum is $1.5\times10^{-33}$. The TS of the flat circle (with log-parabolic spectral shape) with respect to the null hypothesis is 988, which roughly corresponds to a detection significance at the level of 31 standard deviations. The best fit parameters are N$_0=(17.5\pm0.2)10^{-12}$\,cm$^{-2}$s$^{-1}$MeV$^{-1}$, $\alpha=2.596\pm0.013$, $\beta=0.338\pm0.008$, where uncertainties are only statistical. The total energy flux is $(1.10\pm0.03)\times10^{-4}$\,MeV\,cm$^{-2}$s$^{-1}$, with a maximum in the spectral energy distribution at $413\pm11$\,MeV.

To explore possible spectral differences throughout \hb, we divided the circle in three pieces covering $120\arcdeg$ each and let them acquire different values for the $N_0$, $\alpha$ and $\beta$. In this way, segment NE covers clouds A and N; segment NW covers cloud NW and segment S covers cloud S. We used the likelihood ratio $-2\log(\mathcal{L}_{segment}/\mathcal{L}_{circle})$, where $\mathcal{L}_{segment}$ refers to the segmented circle without the tested segment. With this likelihood ratio, two thirds of the circle may be regarded as part of the null hypothesis, thus providing a measure of the significance of the tested segment. All three segments contribute significantly to the overall emission. The likelihood ratio values of each of these segments are summarized in Table~\ref{partcircle}, along with the flux corresponding to each segment and their spectral parameters. The maximum energy flux is attained at $E_{max}=e^\frac{2-\alpha}{2\beta}$\,GeV. To evaluate the uncertainty in $E_{max}$, the uncertainties in $\alpha$ and $\beta$ are taken into account, as well as their covariance. We find a hint that the segment NW has a softer spectrum, peaking at lower energies than the other two regions (see also Fig.~\ref{spec}).

\begin{table*}
  \caption[]{Spectral analysis of the different circle segments above 100\,MeV.}
  \label{partcircle}
  \centering{}
  \begin{tabular}{llllll}
    \hline\hline
    Segment  & LR & Flux [$10^{-8}$cm$^{-2}$s$^{-1}$] ($b$) & $\alpha\, (a)$ & $\beta\, (a)$ & $E_{max}$ [GeV] ($b$) \\
    \hline
    Global     &   0 & $17.5\pm0.2$ & $2.596\pm0.013$ & $0.338\pm0.008$ & $0.413\pm0.011$ \\ 
    North-East &  90 & $4.0\pm0.9$  & $2.41\pm0.16$   & $0.33\pm0.09$   & $0.54\pm0.14$   \\ 
    North-West & 116 & $7.6\pm1.2$  & $2.87\pm0.15$   & $0.32\pm0.07$   & $0.26\pm0.06$   \\
    South      & 151 & $5.3\pm1.0$  & $2.49\pm0.13$   & $0.39\pm0.09$   & $0.53\pm0.10$    \\
    \hline
  \end{tabular}
  \\
($a$) Spectral parameters of the log-parabolic spectral shape (equation~\ref{logparabola}).\\
($b$) Integral flux and the energy at which the energy flux is maximum.
\end{table*}

\section{Discussion}
\label{discuss}
The luminosity between 100\,MeV and 5.6\,GeV is $L=(1.34\pm0.03_{\rm stat})\times10^{34}(d/0.8\mathrm{kpc})^2$\,erg/s. Unless $d=1.7$\,kpc is confirmed, \hb\, belongs to the group of low-luminosity, GeV-emitting SNRs, such as Cygnus Loop \citep{cygloop} or S147 \citep{s147}, which are clearly less luminous than the first GeV-emitting SNRs that were discovered. For instance, W51C \citep{w51}, IC443 \citep{ic443}, W49B \citep{w49b}, or Cas A \citep{casA} have luminosities $L>10^{35}$\,erg/s. Also the break in energy is found at lower energy in \hb\, than in the case of the luminous SNRs.

Gamma-ray emission from supernova remnants can  be produced by several non-exclusive mechanisms. Electrons and positrons in the supernova remnant shell interact by inverse Compton scattering with ambient photon fields (such as infrared starlight) to produce gamma rays. Another possibility is that electron-atom or electron-molecule interactions in a dense medium result in bremsstrahlung radiation. These mechanisms involve electrons accelerated up to the energies of the observed gamma rays. Complementarily, gamma rays can also be produced by neutral pion decay, where $\pi^0$s result from the collision of accelerated protons (or heavier nuclei) with nucleons of the ambient gas. The absence of nonthermal X-ray emission favors a hadronic origin of the observed gamma-ray emission. Moreover, the rapid steepening of the spectrum above few GeV is the kind of signature expected from the re-acceleration of pre-existing cosmic rays \citep{blandford,uchiyama}, where high energy cosmic rays escape from the SNR confinement region \citep{zirakaronian}.

To check the viability of the leptonic and hadronic scenarios from the energetics point of view, we consider the energy from the supernova explosion that is converted into accelerated particles, $W=L\times\tau$. In this expression, $L$ is the gamma-ray luminosity and $\tau$ the characteristic cooling time of the dominant accelerated particle type. When the gamma-ray luminosity is \textit{hadronic}-dominated, $\tau_p\sim5.3\times10^7(n/1\mathrm{cm}^{-3})^{-1}$\,years is the cooling time of the accelerated protons as a function of the ambient proton density $n$ \citep{Aharonian}. According to \cite{koo}, cloud S has density $\sim7000$\,cm$^{-3}$, and central evaporating clouds have densities $\sim4\times10^4$\,cm$^{-3}$. Moreover, \citet{tatematsu} quote a density of about 100\,cm$^{-3}$ for cloud A. We computed our own estimate of the average density of the region. To estimate the total mass we used the CO data from CfA 1.2\,m Millimeter-Wave Telescope \citep{dame}. We assumed a standard linear relationship between the velocity integrated CO intensity, $I_{\rm CO}$, and the molecular hydrogen column density, $ N(\rm H_2)$:
\begin{equation}
  N(\rm H_2)/I_{\rm CO}=(1.8 \pm 0.3) \times 10^{20}\rm cm^{-2}K^{-1}km^{-1}s^{-1}
\end{equation}
as derived by \citet{dame2001}. This equation yields $M_{\rm CO} /M_{\odot} = 1200 S_{\rm CO} d^2_{kpc}$, where $d_{\rm kpc}$ is the distance to the cloud in kpc, and $S_{\rm CO}$ the CO emission integrated over velocity and the angular extent of the cloud in $\rm K~km~s^{-1}~arcdegree^2$. We conclude that there are as many as 12000 solar masses of molecular gas coinciding with the gamma-ray emission in the velocity range from -20 to 0\,km/s\footnote{\citet{byun} show that the majority of the emission is concentrated in the range -20 to 0 km/s. Although there is some emission at higher velocities, at about 3\,km~s$^{-1}$, telluric CO emission corrupts the spectra. To avoid this issue, and for simplicity, we took the velocity range -20 to 0 km/s.}. Assuming that this gas is in a spherical volume of 25\,pc diameter we obtain an average density of about 60 protons per cubic centimeter, and therefore $W_p\sim4\times10^{47}$\,erg. Instead, in case the gamma-ray luminosity is \textit{leptonic}-dominated, the cooling time due to bremsstrahlung interactions depends on the ambient density as $\tau_{brems}\sim4\times10^7(n/\rm{cm}^3)^{-1}$ \citep{Aharonian}, which leads to a live time $\tau_{brems}\sim7\times10^{5}$ years and requires an energy budget similar to the hadronic interactions, $W_e\sim3\times10^{47}$\,erg. Alternatively, the cooling time of electrons due to synchrotron losses is $\tau_{sync}\sim1.3\times10^{10}(B/1\mu\mathrm{G})^{-2}(E_{e^-}/1\mathrm{GeV})^{-1}$\,years \citep{gaisser}. Therefore, an $E_{e^-}=3$\,GeV electron in a typical interstellar magnetic field of 3\,$\mu$G would have a live time of $\tau_{sync}\sim5\times10^8$ years. Only with a magnetic field as high as 80\,$\mu$G, could the synchrotron losses become efficient enough to make them comparable to the losses due to bremsstrahlung interactions. Although detailed modeling is needed to confirm these considerations, our order-of-magnitude estimates favor the view that the high-density regions surrounding the SNR determine the dominant emission process. This is supported further by the coincidence of the gamma-ray bright regions with dense clumps of molecular gas (Fig.~\ref{signimaps}). In any case, energy budget considerations are not sufficient to distinguish between electron-dominated and proton-dominated scenarios under the assumption that the supernova explosion has a typical energy release of about $10^{51}$\,erg. 

Northeast of the circle there is a spot that becomes bright with increasing energy (Fig.~\ref{signimaps}). We discuss the possibility that this comes from a hard gamma-ray emission component related to cloud A. This cloud most likely already existed at the time of the SNR explosion, it is not shocked, and its velocity is not significantly different from that of the wall \citep{tatematsu}. According to \citet{koo}, this cloud does not present broad line emission, so there is no direct evidence of interaction with \hb\footnote{The lack of broad line emission is also the case for 3c391, where the interaction with molecular clouds is certain \citep{ReachRho2002}.}. Moreover, cloud A coincides with a concavity of the SNR shell. The coincidence of cloud A with this feature in the radio continuum emission suggests that the overtaking of this cloud leads to retardation of the shock front in comparison to the surrounding, although high-resolution CO maps from \cite{koo} did not find such evidence. There is still the possibility that the shock is dissociative, and molecules have not been reformed (thus being omitted in the search of broad line emission regions), but the shock velocity of 20\,km~s$^{-1}$ observed by Koo is in principle not enough for molecule dissociation (which typically requires 25--50\,km~s$^{-1}$). Finally, the infrared emission of cloud A is caused by dust heated by the ambient radiation fields in their surface, whereas in clouds N and S it is due to lines from shock-excited molecules \citep{koo}. In fact, \cite{koo} mentions a diffuse component connecting clouds N and A. All in all, there is evidence that cloud A is different to other molecular clouds in the vicinity of \hb. As mentioned in section~\ref{intro} there is also the suggestion that cloud A could be in the foreground of \hb, related to the Cyg OB7 association (i.e. at 0.8\,kpc), whereas \hb\, could be in the background, at 1.7\,kpc or more. If we assume that the gamma-ray brightness of cloud A is due to runaway protons from \hb, we can estimate the maximum distance between the two objects by the relation $R_{d}=\sqrt{4Dt}$, where $D\sim10^{28}$\,cm$^2$s$^{-1}$ \citep[][equation 11]{Gabici2009} is the diffusion coefficient of cosmic rays protons of 10\,GeV (which originate 1\,GeV gamma rays). In this case the separation between cloud A and \hb\, would be roughly 50\,pc. We note that $D$ is completely unknown, and this distance is not to be taken as a measure of the separation between the cloud and the SNR, but only a suggestion that cloud A is indeed close to the SNR. We cannot conclude the same about the two other clouds (B and C) on the eastern rim detected by \cite{tatematsu}. These are dark in gamma rays and therefore they may be well in the foreground, as suggested by \citet{byun}. 

Finally, we consider it interesting that the part of the circle related to cloud NW presents a spectral break at lower energies than the rest of the SNR. From the $E>3$\,GeV map (Fig.~\ref{3GeV}), we see that the majority of the emission in this region comes from a bow-shaped structure that resembles the SNR shell itself. We suggest that the gamma-ray emission from this region originates in the shell itself, or from a molecular cloud (probably cloud NW) that was overtaken at an earlier stage than those producing the emission in the remaining two thirds of the remnant.

\section{Conclusions}
The analysis of 3.5 years of public \fermi\, data leads to a clear detection of an extended source of gamma rays coincident with the SNR \hb. Our own method for visualizing the morphology allows us to describe the source as a flat circle with uniform emission, although we resolve a clumpy structure at the highest energies. The spectral analysis reveals a peak of the gamma-ray emission at $413\pm11$\,MeV. However, we conduct a dedicated spectral analysis for three regions, which show particularly bright emission above 500\,MeV coincident with dense molecular hydrogen clumps. We find indications that the gamma-ray emission peaks at somewhat lower energies ($0.26\pm0.06$\,GeV) in the NW region of the shell (coincident with cloud NW and the shell itself), whereas in the NE (coincident with clouds N and A) and the S regions (coincident with cloud S) the peak is found around $\sim 0.5$\,GeV. 

Such spectral breaks are expected in middle-aged SNRs like \hb\, due to the escape of CRs from the confinement region. In \hb\, the break occurs at lower energies than in other similar objects. This could be related to the proximity of very dense molecular clouds to the supernova explosion progenitor, which could have slowed down the shock rapidly \citep{ohira}. The spectrum from the region related to the bow-shaped emission in the NW peaks at lower energies than the spectrum from clouds N and S. This fact matches the understanding that the most energetic particles related to the SNR may have already escaped the NW region, due to a smaller confinement volume, or because cloud NW was shocked at an earlier stage than clouds N or S. Moreover, in the S/N map above 3\,GeV, we see how a spot coincident with cloud A that becomes relatively brighter than the other clouds with increasing energy. We suggest that this is because cloud A is separated from the SNR, and it is now reproducing the spectrum of the SNR at an earlier stage due to particles that diffused away from the shock \citep{ahatoyan}.

Therefore, given that \hb\, appears as an extended object even for gamma-ray telescopes, it provides the opportunity to observe the production and diffusion of accelerated particles (most likely protons), from the SNR shell to distant molecular clouds acting as targets.

\begin{acknowledgements}
  This work has been funded by projects DE2009-0064 and FPA2009-07474 from the Spanish Ministry of Research, Development and Innovation (former MICINN).

  We are thankful to professor Felix Aharonian for the very useful discussion and suggestions.

\end{acknowledgements}

\bibliography{HB21}
\bibliographystyle{aa}

\end{document}